\documentclass[10pt,pra,twocolumn,floatfix]{revtex4}
\usepackage{amsfonts}
\usepackage{amssymb}
\usepackage{amsmath}
\usepackage[dvips]{graphicx}
\usepackage{color}

\begin{document}

\title{Experimental test of error-tradeoff uncertainty relation using a
continuous-variable entangled state}
\author{Yang Liu$^{1,2}$}
\author{Zhihao Ma$^{3}$}
\author{Haijun Kang$^{1,2}$}
\author{Dongmei Han$^{1,2}$}
\author{Meihong Wang$^{1,2}$}
\author{Zhongzhong Qin$^{1,2}$}
\author{Xiaolong~Su$^{1,2}$}
\email{suxl@sxu.edu.cn}
\author{Kunchi Peng$^{1,2}$}

\affiliation{$^1$State Key Laboratory of Quantum Optics and Quantum Optics Devices,
Institute of Opto-Electronics, Shanxi University, Taiyuan 030006, China\\
$^2$Collaborative Innovation Center of Extreme Optics, Shanxi University,
Taiyuan, Shanxi 030006, China\\
$^3$Department of Mathematics, Shanghai Jiaotong University, Shanghai,
200240, China\\
}

\begin{abstract}
Heisenberg's original uncertainty relation is related to measurement
effect, which is different from the preparation uncertainty
relation. However, it has been shown that Heisenberg's
error-disturbance uncertainty relation can be violated in some cases.
We experimentally test the error-tradeoff uncertainty relation by
using a continuous-variable Einstein-Podolsky-Rosen (EPR) entangled
state. Based on the quantum correlation between the two entangled
optical beams, the errors on amplitude and phase quadratures of one
EPR optical beam coming from joint measurement are estimated
respectively, which are used to verify the error-tradeoff relation.
Especially, the error-tradeoff relation for error-free measurement
of one observable is verified in our experiment. We also verify the
error-tradeoff relations for nonzero errors and mixed state by
introducing loss on one EPR beam. Our experimental results
demonstrate that Heisenberg's error-tradeoff uncertainty relation is
violated in some cases for a continuous-variable system, while the
Ozawa's and Brainciard's relations are valid.
\end{abstract}

\maketitle

\section{ Introduction}

As one of the cornerstones of quantum mechanics, uncertainty relation
describes the measurement limitation on two incompatible observables \cite%
{Heisenberg}. It should be emphasized that the uncertainty relation actually
states an intrinsic property of a quantum system, rather than a statement
about the observational success of current technology. Uncertainty relation
has deep connection with many special characters in quantum mechanics, such
as Bell non-locality and entanglement \cite{Bell2,EPR2}, which cannot occur
in classical world. With rapid progress in quantum technology, such as
quantum communication and quantum computation \cite{COMMUN,FURU}, in recent
years, it is important for us to know the fundamental limitations in the
achievable accuracy of quantum measurement.

Note that there are two different types of uncertainty relations, one is the
preparation uncertainty relation, which studies the minimal dispersion of
two quantum observables before measurement \cite{Kennard,Rob29}. The
Robertson uncertainty relation \cite{Rob29}, reads as $\sigma (x)\sigma
(p)\geq \hbar /2$, is a typical example in this sense, where $\sigma (x)$
and $\sigma (p)$ are the standard deviations of position and momentum of a
particle. For such uncertainty relation, the measurements of $x$ and $p$ are
performed on an ensemble of identically prepared quantum systems. While in
the original spirit of Heisenberg's idea \cite{Heisenberg}, the Heisenberg's
uncertainty principle should be based on the observer's effect, which means
that measurement of a certain system cannot be made without affecting the
system. So this leads to the second type of uncertainty relation:
measurement uncertainty relation, which studies to what extent the accuracy
of position measurement of a particle is related to the disturbance of the
particle's momentum, so called the error-disturbance uncertainty relation
\cite{Ozawa03}. It is also called the error-tradeoff relation in the
approximate joint measurements of two incompatible observables \cite%
{Ozawa04,Branciard}.

Heisenberg's error-tradeoff uncertainty relation for joint measurement is generally
expressed as%
\begin{equation}
\varepsilon (A)\varepsilon (B)\geq C_{AB}  \label{H}
\end{equation}%
where $C_{AB}=\left\vert \left\langle [A,B]\right\rangle \right\vert /2$, $%
[A,B]=AB-BA$. However, it has been shown that this relation is not valid in
some cases \cite{Balllentine}. For this reason, Ozawa and Hall proposed new
measurement uncertainty relations which have been theoretically proven to be
universally valid for any incompatible observables, respectively \cite%
{Ozawa03,Ozawa04,Hall04}. After that, Branciard proposed a new uncertainty
relation, which is universally valid and tighter than the Ozawa's relation
\cite{Branciard}. There are also other types of measurement uncertainty
relations generalizing Heisenberg's original idea, which can be found in
Refs. \cite%
{Werner2,PhysRevA022106,PhysRevA032,PhysRevLett050401,lu,Barchielli2017}.
Experimental tests of the measurement uncertainty relations have been
demonstrated in photonic \cite%
{EXPphotons1,EXPphotons2,EXPphotons3,EXPphotons4,EXPphotons5,EXPphotons6},
spin \cite{EXPpolarizedneutrons1,EXPphotons7,EXPphotons8,EXPW1}, and ion
trap systems \cite{EXPW2}. All of these experiments are limited in
discrete-variable systems. Up to now, experimental test of the measurement
uncertainty relation based on continuous-variable system has not been
reported.
\begin{figure}[tbp]
\begin{center}
\includegraphics[width=80mm]{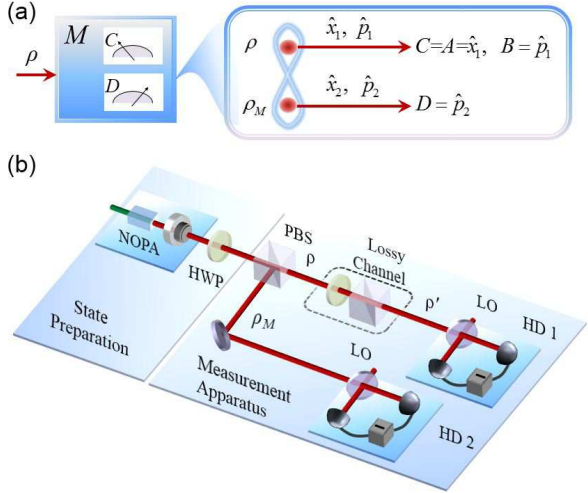}
\end{center}
\caption{ (a) Schematic of the test principle for
error-tradeoff relation by joint measurement on a continuous-variable
entangled state. A quantum state $\protect\rho $ is measured in a joint
measurement apparatus $M$, where two compatible observables $C$ and $D$ are
measured simultaneously to approximate two incompatible observables $A$ and $%
B$, respectively. The right inset describes the joint measurement apparatus
for the error-free measurement of observable $A.$ (b) Schematic of
experimental setup. An EPR entangled state is produced by a NOPA operating
in the state of deamplification. The two modes of EPR state are used as the
signal state $\protect\rho $ and the meter state $\protect\rho _{M}$ which
are detected by the homodyne detectors HD1 and HD2, respectively. The lossy
channel is simulated by a half-wave plate (HWP) and a polarization beam
splitter (PBS).\ LO: local oscillator.}
\end{figure}

In this paper, we present the first experimental test of the error-tradeoff relation
for two incompatible variables, amplitude and phase quadratures of an
optical mode, using a continuous-variable EPR entangled state. Based on
quantum correlations of the EPR entangled beams, the error-tradeoff relation
with zero error (error-free) of one observable is verified directly by
performing joint measurement on two EPR beams. In this case, Heisenberg's error-tradeoff uncertainty
relation is violated, while Ozawa's and Branciard's relations are valid. We
also test the error-tradeoff relations for nonzero errors and mixed state by
introducing loss on signal mode. Our experimental test of the
continuous-variable error-tradeoff relations makes the test of the
measurement uncertainty relation more complete.

\section{Theoretical framework}

One mode of EPR entangled state is used as signal state $\rho $ and two
incompatible observables are taken as $A=\hat{x}_{1}$ and $B=\hat{p}_{1}$,
respectively [Fig. 1(a)], where $\hat{x}_{1}=(\hat{a}+\hat{a}^{\dag })/2$ and $%
\hat{p}_{1}=(\hat{a}-\hat{a}^{\dag })/2i$ denote the amplitude and phase
quadratures of $\rho $, respectively. Another mode of EPR entangled state is
used as the meter state $\rho _{M}$. Two compatible observables $C$ and $D$
are measured simultaneously to approximate $A$ and $B.$ The quality of the
approximations are characterized by defining the root-mean-square errors $%
\varepsilon (A)=\langle (C-A)^{2}\rangle ^{1/2}$ and $\varepsilon
(B)=\langle (D-B)^{2}\rangle ^{1/2}$. Ozawa's error-tradeoff relation is
expressed by \cite{Ozawa03,Ozawa04}
\begin{equation}
\varepsilon (A)\varepsilon (B)+\varepsilon (A)\sigma (B)+\sigma
(A)\varepsilon (B)\geqslant C_{AB}  \label{O}
\end{equation}%
where $\sigma (A)$\ is the standard deviation of observable $A$. The
Branciard's error-tradeoff relation is given by \cite{Branciard}%
\begin{eqnarray}
&&\big[\varepsilon ^{2}(A)\sigma ^{2}(B)+\sigma ^{2}(A)\varepsilon ^{2}(B)
\notag  \label{B} \\
&&+2\varepsilon (A)\varepsilon (B)\sqrt{\sigma ^{2}(A)\sigma
^{2}(B)-C_{AB}^{2}}\big]^{1/2}\geqslant C_{AB}
\end{eqnarray}%
where the parameter $C_{AB}=1/4$ denote that $A$ and $B$ cannot be jointly
measured on $\rho $ simultaneously. The variances of the amplitude and phase
quadratures of two EPR beams are expressed as $\sigma ^{2}(\hat{x}%
_{1})=\sigma ^{2}(\hat{p}_{1})=\sigma ^{2}(\hat{x}_{2})=\sigma ^{2}(\hat{p}%
_{2})=(e^{2r}+e^{-2r})/8$, where $r$ is the squeezing parameter \cite{FURU}.
In the experiment, we test Heisenberg's, Ozawa's and Branciard's error-tradeoff uncertainty relations
in three cases, i.e., error-free measurement of one observable, nonzero
error and mixed state cases.

\section{Experimental implementation and results}

In the experiment, an EPR entangled state with $-$2.9 dB squeezing
and 3.9 dB antisqueezing is prepared by a nondegenerate optical
parametric amplifier (NOPA), as shown in Fig. 1(b), which consists
of an $a$-cut type-II KTP crystal and a concave mirror \cite{EPRSU}.
The front face of the KTP crystal is used as the input coupler, and
the concave mirror with 50 mm curvature serves as the output
coupler. The front face of the KTP crystal is coated with the
transmission of 42\% at 540 nm and high reflectivity at 1080 nm. The
end face of the KTP crystal is antireflection coated for both 540 nm
and 1080 nm. In the measurement, a sample size of 5$\times
10^{5\text{ }}$data points is used for all quadrature measurements
with sampling rate of 500 K/s. The interference efficiency between
signal and local oscillatior is 99\% and the quantum efficiency of
photodiodes are 99.6\%.

At first, we consider a situation that the observable $A$ is measured
accurately (error-free measurement of observable $A$), i.e., the optimal
estimation $C=A$. The measured phase quadrature $D=\hat{p}_{2}$ is used to
approximate the observable $B$. Because the amplitude quadrature $\hat{x}_{1}
$ of $\rho $ and the phase quadrature $\hat{p}_{2}$ of $\rho _{M}$ are
compatible, they can be measured simultaneously. The errors for
approximating $A$ and $B$ are expressed as $\varepsilon (A)=\sqrt{\langle
(C-A)^{2}\rangle }=0$, and $\varepsilon (B)=\sqrt{\langle (D-B)^{2}\rangle }=%
\sqrt{\sigma ^{2}(\hat{p}_{2}-\hat{p}_{1})}=e^{-r}/\sqrt{2}$, respectively.
Since $\varepsilon (A)=0$ and $\varepsilon (B)<\infty $, we have
\begin{equation}
\varepsilon (A)\varepsilon (B)=0.
\end{equation}%
It is obvious that Heisenberg's error-tradeoff uncertainty relation is violated. The Ozawa's and
Branciard's relations are the same for $\varepsilon (A)=0$, which are
\begin{equation}
\sigma (A)\varepsilon (B)=\sqrt{1+e^{-4r}}/4\geqslant 1/4.
\label{error-free}
\end{equation}

The amplitude quadrature $\hat{x}_{1}$ of the signal state is measured by a
homodyne detector HD1 in the time domain, as shown in Fig. 1(b). To evaluate
the error $\varepsilon (B)$, we experimentally measure the observables $B$
and $D$, i.e. the phase quadratures $\hat{p}_{1}$ and $\hat{p}_{2}$, by two
homodyne detectors (HD1 and HD2) simultaneously.

\begin{figure}[tbp]
\begin{center}
\includegraphics[width=80mm]{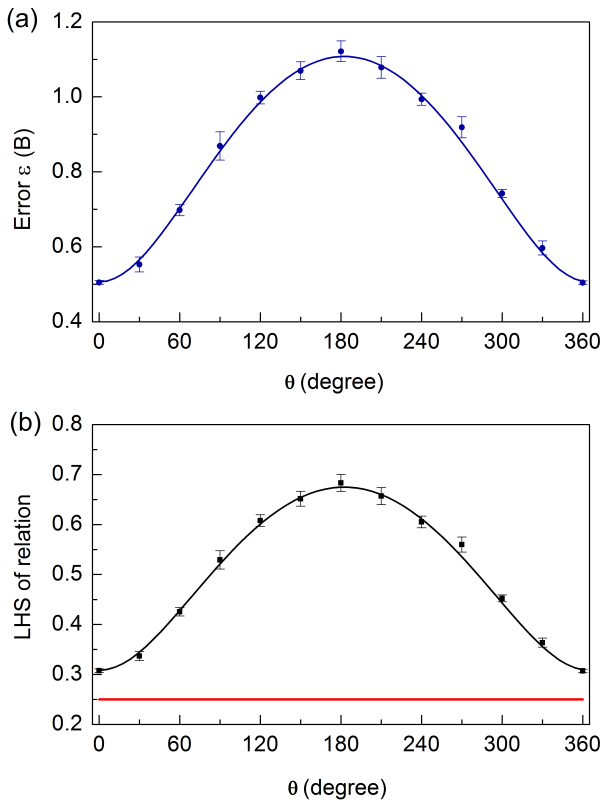}
\end{center}
\caption{Results of the uncertainty relation in case of
error-free measurement of observable $A$. (a) The error $\protect%
\varepsilon (B)$ as a function of the relative phase. (b) The LHS of
the Ozawa's and Branciard's relation as a function of the relative phase.
The right hand side of the relations $C_{AB}$ is indicated by the red line.
The solid curves and data points are the theoretical calculated result and
experimental results, respectively. The error bars are obtained by
root-mean-square of repeated measurements for ten times. The experimentally
measured results are in good agreement with the theoretical calculation.}
\end{figure}

\begin{figure}[tbp]
\begin{center}
\includegraphics[width=80mm]{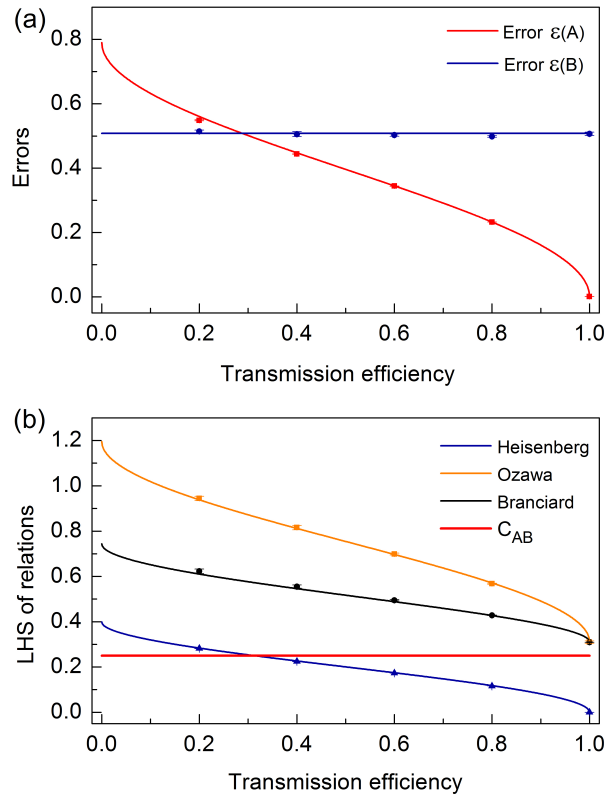}
\end{center}
\caption{Results of the uncertainty relations in case of
nonzero errors. (a) The errors $\protect\varepsilon (A)$ (red curve)
and $\protect\varepsilon (B)$ (blue curve) as functions of the transmission
efficiency. (b) The LHS of the relations as functions of the
transmission efficiency. Blue curve: the Heisenberg's relation in Eq. (1).
Yellow curve: the Ozawa's relation in Eq. (2). Black curve: the Branciard's
relation in Eq. (3). The right hand side of the relations $C_{AB}$ is
indicated by the red line. }
\end{figure}

In our experiment, the achievable lower bound is limited by the quantum
correlation of the EPR entangled state [Eq.~(\ref{error-free})]. In order to
demonstrate this property, we change the quantum correlation of signal state
and meter state by changing the relative phase $\theta $ between the two
mode of EPR entangled state. Thus, the error $\varepsilon (B)=\sqrt{\sigma
^{2}(e^{i\theta }\hat{p}_{2}-\hat{p}_{1})}$ is measured in experiment. When
the relative phase $\theta =0^{\circ }$ and $\theta =360^{\circ }$, the
minimum error is obtained [Fig. 2(a)] and the left-hand-side (LHS) of the
relation reaches its minimum value [Fig. 2(b)], which is determined by the
present squeezing level. When$\ \theta =180^{\circ }$, the maximum error is
obtained, which corresponds to the measurement of anti-correlated noise $%
\sqrt{\sigma ^{2}(\hat{p}_{2}+\hat{p}_{1})}$. The results confirm that the
Ozawa's and Branciard's relations are the same and valid for the error-free
measurement of observable $A$.
\begin{figure}[tbp]
\begin{center}
\includegraphics[width=80mm]{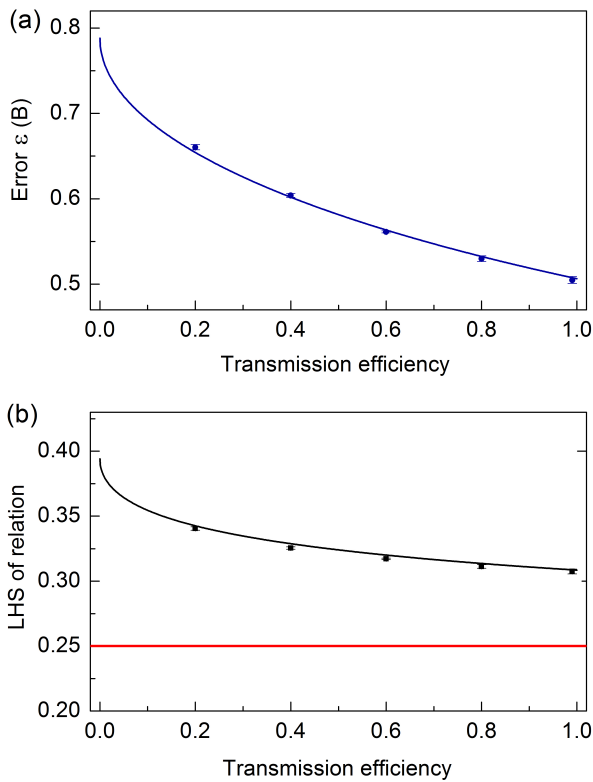}
\end{center}
\caption{ Uncertainty relation for mixed state. (a)
The error $\protect\varepsilon (B)$ as a function of the transmission
efficiency. (b) The LHS of the Ozawa's and Branciard's relation as a
function of the transmission efficiency. The right hand side of the
relations $C_{AB}$ is indicated by the red line. }
\end{figure}

Then, we test the error-tradeoff relation with nonzero errors. When both
errors are not equal to zero, Ozawa's and Branciard's relations are
different. In the experiment, we apply a linear operation on the signal
mode, which is done by transmitting the signal mode through a lossy channel,
as shown in the inset of Fig. 1(b). In this case, the amplitude and phase
quadratures of the signal mode are changed to $\hat{x}_{1}^{^{\prime }}=%
\sqrt{T}\hat{x}_{1}+\sqrt{1-T}\hat{x}_{v}$ and $\hat{p}_{1}^{^{\prime }}=%
\sqrt{T}\hat{p}_{1}+\sqrt{1-T}\hat{p}_{v}$, respectively, after transmitted
over the lossy channel, where $\hat{x}_{v}$ and $\hat{p}_{v}$\ represent the
amplitude and phase quadratures of vacuum. By choosing $C=\hat{x}%
_{1}^{^{\prime }}$ and $D=\hat{p}_{2}$, which are compatible, the errors for
the two incompatible observables $A=\hat{x}_{1}$ and $B=\hat{p}_{1}$ are $%
\varepsilon (A)=\sqrt{\sigma ^{2}(\hat{x}_{1}^{^{\prime }}-\hat{x}_{1})}$
and $\varepsilon (B)=\sqrt{\sigma ^{2}(\hat{p}_{2}-\hat{p}_{1})}$,
respectively.

In this case, the error $\varepsilon (A)$ increases with the decreasing of
channel efficiency, while the error $\varepsilon (B)$ is not affected by the
channel efficiency [Fig. 3(a)]. Heisenberg's error-tradeoff unceratinty relation is violated when the
transmission efficiency is higher than 0.3. While the Ozawa's and
Branciard's relations are always valid [Fig. 3(b)]. By comparing the LHS of
Ozawa's and Branciard's relation, we confirm that Branciard's relation is
tighter than Ozawa's relation.
\begin{figure}[tbp]
\begin{center}
\includegraphics[width=80mm]{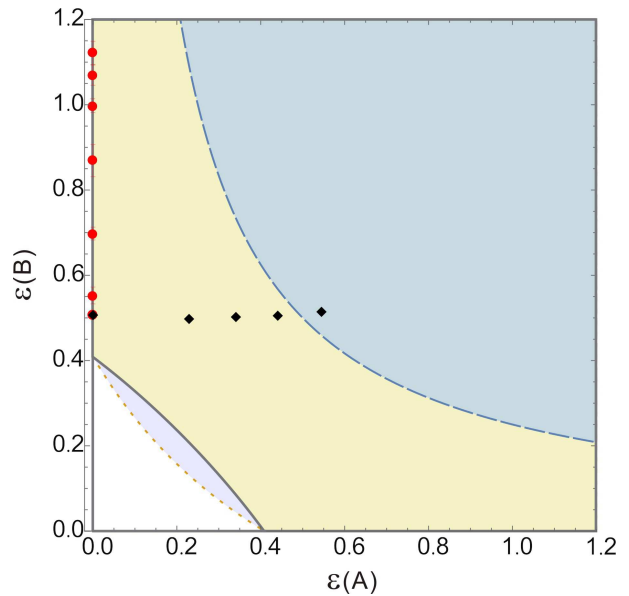}
\end{center}
\caption{Lower bounds of the error-tradeoff relations. Blue
dashed curve: the Heisenberg's bound. Yellow dotted curve: the Ozawa's
bound. Gray solid curve: the Branciard's bound. Red circles: experimental
data for error free measurement of observable $A$ as shown in Fig. 2. Black
diamonds: experimental data for nonzero errors condition as shown in Fig. 3.
}
\end{figure}

Finally, we demonstrate the error-tradeoff relation for mixed state, i.e.,
the state $\rho $ transmitted over a lossy channel. Here, observables $C=A=%
\hat{x}_{1}^{^{\prime }}$, $B=\hat{p}_{1}^{^{\prime }}$, and $D=\hat{p}_{2}$
are chosen, and thus errors for the mixed state are $\varepsilon (A)=0$ and $%
\varepsilon (B)=\sqrt{\sigma ^{2}(\hat{p}_{2}-\hat{p}_{1}^{^{\prime }})}$,
respectively. In this case, Ozawa's and Branciard's relations are the same.
The error $\varepsilon (B)$ and the LHS of the relation increase along with
the decreasing of transmission efficiency as shown in Fig. 4(a) and 4(b),
respectively. The error and LHS of the relation get the minimum value when
the transmission efficiency is unit.

The predicted lower bounds for Heisenberg's [Eq. (\ref{H})], Ozawa's [ Eq. (%
\ref{O})] and Brinciard's [Eq. (\ref{B})] error-tradeoff relations are
compared in the plane ($\varepsilon (A),\varepsilon (B)$), as shown in Fig.
5. For the Heisenberg's error-tradeoff uncertainty relation (bounded by the blue dashed curve), one of
the error must be infinite when the other goes to zero. While in our
experiment, for the case of error $\varepsilon (A)=0$, the finite error $%
\varepsilon (B)$ is observed (red circles), which violates the
Heisenberg's error-tradeoff uncertainty relation, yet satisfies the
Ozawa's and Branciard's relation. For the case of nonzero errors,
only one of the observed values satisfies the Heisenberg's
error-tradeoff uncertainty relation (the data with 0.2 transmission
efficiency). Our experimental data do not reach the lower bound of
the relations for the limitation of the experiment condition, for
example the limited squeezing parameter.

\section{Conclusion}

We experimentally test the Heisenberg's, Ozawa's and Branciard's
error-tradeoff relations for continuous-variable observables, i.e.,
amplitude and phase quadratures of an optical mode. Especially, we
investigate the error-tradeoff relation in case of zero error by using
Gaussian EPR entangled state. Three different measurement apparatus are
applied in our experiment, which are used to test the error-tradeoff
relation for three different cases. The results demonstrate that the
Heisenberg's error-tradeoff uncertainty relation is violated in some cases while the Ozawa's and the
Brinciard's relations are valid. Our work is useful not only in understanding
fundamentals of physical measurement but also in developing of continuous
variable quantum information technology.

\section*{ACKNOWLEDGMENTS}

This research was supported by the NSFC (Grant Nos. 11834010,and 61601270), the program of Youth Sanjin Scholar, the Applied
Basic Research Program of Shanxi Province (Grant No. 201601D202006),
National Basic Research Program of China (Grant No. 2016YFA0301402), and the
Fund for Shanxi ``1331 Project" Key Subjects Construction.

Y.L. and Z.M. contributed equally to this work.

\end{document}